\documentclass[aps,prl,twocolumn,superscriptaddress,showpacs]{revtex4}

\usepackage{graphicx}

\begin{document}

\newcommand{\be}{\begin{equation}}
\newcommand{\ee}{\end{equation}}
\newcommand{\bea}{\begin{eqnarray}}
\newcommand{\eea}{\end{eqnarray}}

\title{Magnetization noise in magnetoelectronic nanostructures}

\author{J\o rn Foros}\affiliation{Department of Physics, Norwegian University of Science and Technology, 7491 Trondheim, Norway}
\author{Arne Brataas}\affiliation{Department of Physics, Norwegian University of Science and Technology, 7491 Trondheim, Norway}
\author{Yaroslav Tserkovnyak}\affiliation{Lyman Laboratory of Physics, Harvard University, Cambridge, Massachusetts 02138, USA}
\author{Gerrit E. W. Bauer}\affiliation{Kavli Institute of NanoScience, Delft University of Technology, 2628 CJ Delft, The Netherlands}

\date{\today}

\begin{abstract}By scattering theory we show that spin current noise in normal electric conductors in contact with nanoscale ferromagnets increases the magnetization noise by means of a fluctuating spin-transfer torque. Johnson-Nyquist noise in the spin current is related to the increased Gilbert damping due to spin pumping, in accordance with the fluctuation-dissipation theorem. Spin current shot noise in the presence of an applied bias is the dominant contribution to the magnetization noise at low temperatures.

\end{abstract}


\maketitle

Time-dependent fluctuations of observables (``noise") are a nuisance for the engineer but also a fascinating subject of study for the physicist. The thermal current fluctuations in electric circuits as well as the Poissonian current fluctuations due to the discrete electron charge emitted by hot cathodes are classical textbook subjects. The fluctuations of the order parameter in ferromagnets, such as Barkhausen noise due to moving domain walls, have been studied by the magnetism community for almost a century. Recently, it has been discovered that electronic noise is dramatically modified in nanostructures. Theoretical predictions on the suppression of charge shot noise in quantum devices have been confirmed experimentally \cite{Blanter-review}. Spin current fluctuations, i.e., spin shot noise, is as yet a purely theoretical concept \cite{Nazarov-review}. In nanoscale magnetism, thermal noise plays an important role by activating magnetization reversal of ferromagnetic clusters \cite{WernsdorferKoch}. Charge shot noise in ferromagnetic spin valve devices has been discussed as well \cite{Tserkovnyak-prb2001,FNF-SN}. Interesting new questions have been raised by recent experimental studies on the dynamics of nano-scale spin valves \cite{Kiselev,Rippard,Covington} in which electric transport is affected by the magnetization direction of the ferromagnetic elements. Central to these studies is the spin-transfer torque exerted by a spin-polarized current on the magnetization causing it to precess or even reverse direction \cite{Berger,Slonczewski1,MyersKatine}. Covington \emph{et al.} \cite{Covington} interpreted the observed dependence of noise spectra in nano-pillar spin valves on bias current direction in terms of this spin torque, but a full consensus has not yet been reached \cite{Rebei}.

In a normal metal the average current of net spin angular momentum (spin current) vanishes but its fluctuations are finite. In this Letter we demonstrate that equilibrium and non-equilibrium spin current noise in normal metals is directly observable in hybrid ferromagnet-normal metal structures: The noise exerts a fluctuating spin-transfer torque on the magnetization vector causing an observable magnetization noise. The theory of noise in magnetoelectronic devices requires a consistent treatment of fluctuations in the currents as well as the magnetization. We demonstrate that thermal spin current fluctuations are instrumental for the spin pumping-enhanced Gilbert damping in magnetic multilayers \cite{prl88} and that spin shot noise should be observable at low temperatures. The better understanding of noise in ferromagnetic spin valves should aid the development of next-generation magnetoelectronic and magnetic memory devices.

The magnetization noise in isolated single-domain ferromagnets is well described by the Landau-Lifshitz-Gilbert (LLG) equation of motion 
\be	
	\frac{d\mathbf{m}}{dt} = -\gamma\mathbf{m}\times[\mathbf{H}_{\rm eff}+\mathbf{h}^{(0)}(t)] + \alpha_0\mathbf{m}\times\frac{d\mathbf{m}}{dt}, 
\label{LLG} 
\ee
where $m$ is the unit magnetization vector, $\gamma$ the gyromagnetic ratio, $\mathbf{H}_{\rm eff}$ the effective magnetic field, and $\alpha_0$ the Gilbert damping constant. The stochastic torque $\mathbf{m}\times\mathbf{h}^{(0)}(t)$ describes thermal agitation in terms of a random field $\mathbf{h}^{(0)}(t)$ with zero average and a white noise correlation function \cite{Brown} 
\be
	\langle h_{i}^{(0)}(t)h_{j}^{(0)}(t') \rangle = 2k_BT\frac{\alpha_0}{\gamma M_s\mathcal{V}}\delta_{ij}\delta(t-t').
\label{h0corr}
\ee
Here $i$ and $j$ are Cartesian components, $k_BT$ the thermal energy, $M_s$ the saturation magnetization, and $\mathcal{V}$ the volume of the ferromagnet. The magnetization noise depends on the Gilbert damping $\alpha_0$ that parametrizes the dissipation of magnetic energy in the ferromagnet. The relation between noise and damping is a corollary of the fluctuation-dissipation theorem (FDT) \cite{Brown}.

In ferromagnets in contact with normal conductors fluctuating spin currents contribute to the magnetization noise through the spin-transfer torque. The torque is caused by the absorption of only that component of the spin current that is polarized transverse to the magnetization. This happens on the length scale of the magnetic coherence length $\lambda_c = \pi/(k_{F\uparrow}-k_{F\downarrow})$, where $k_{F\uparrow}$ and $k_{F\downarrow}$ are the minority and majority spin Fermi wave numbers in the ferromagnet \cite{prl84epjb22,Waintal,Stiles}. In transition metals $\lambda_c$ amounts to only a couple of monolayers. A second ingredient needed to understand the noise properties is the inverse effect of the spin torque, often referred to as "spin pumping" \cite{Berger,prl88}: a moving ferromagnet in contact with conductors emits a spin current. The loss of angular momentum is equivalent to an enhancement of the Gilbert damping constant such that $\alpha_0\rightarrow\alpha_0+\alpha'$ \cite{prl88}. There is ample evidence that the enhancement $\alpha'$, to be explicitly defined later, can become much larger than $\alpha_0$ \cite{Tserkovnyakreview}.

We consider hybrid structures of a ferromagnet (F) in good electric contact with normal metals (N), such as an N|F|N structure (Fig. 1), with an applied current or voltage bias (a lateral structure in which the ferromagnet is on top of the current carrying normal metal would also serve to illustrate our ideas). At non-zero temperatures the (spin) current through the interface(s), and thus the spin torque, fluctuates. When a bias is applied, the spin current fluctuates even at zero temperature giving spin shot noise. We show in the following that the fluctuations of the magnetization vector due to thermal and shot noise can be described by an effective random field $\mathbf{h}(t)$. The thermal magnetization noise is governed by the FDT, i.e., the relation between the noise amplitude and the Gilbert damping is preserved, with the damping constant $\alpha_0\rightarrow \alpha_0+\alpha'$. In other words, the thermal spin current noise is identified as the microscopic process that ensures validity of the FDT in the presence of spin pumping.

\begin{figure}
\includegraphics{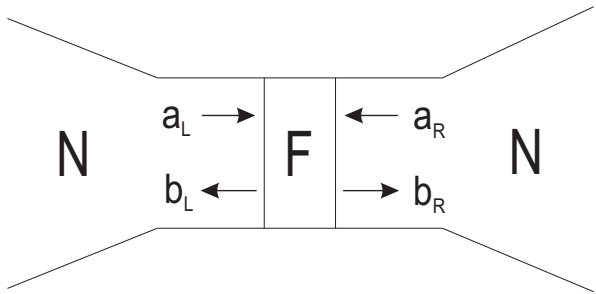}
\caption{\label{system} The transport properties of a thin ferromagnet sandwiched between two large normal metals are evaluated using annihilation and creation operators (only annihilation operators are shown here). $a_{L(R)}$ and $b_{L(R)}$ annihilate an incoming and outgoing electron in the left (right) lead respectively.}
\end{figure}
We use the Landauer-B{\"{u}}ttiker scattering approach \cite{Blanter-review}, generalized to describe spin transport \cite{Tserkovnyak-prb2001} for a thin ferromagnetic film sandwiched by normal metals (Fig. \ref{system}). The normal metals are characterized by chemical potentials $\mu_L$ (left) and $\mu_R$ (right) and a common temperature $T$. The $2\times 2$ distribution matrices $\hat{f}_L$ and $\hat{f}_R$ in spin space reduce to the diagonal form $\hat{f}_L=f(E-\mu_L)\hat{1}$ and $\hat{f}_R=f(E-\mu_R)\hat{1}$, where $f$ is the Fermi-Dirac distribution. The $\alpha\beta$-component of the $2\times 2$ current operator in spin space at time $t$ on the left side of the ferromagnetic film then reads \cite{Tserkovnyak-prb2001}
\bea	
	\hat{I}_{L}^{\alpha\beta}(t) & = & \frac{e}{h}\sum_n\int dEdE'e^{i(E-E')t/\hbar} \\		
				&& \times[a_{Ln\beta}^{\dagger}(E)a_{Ln\alpha}(E')-b_{Ln\beta}^{\dagger}(E)b_{Ln\alpha}(E')]. \nonumber
\eea
Here $a_{Ln\alpha(\beta)}^{(\dagger)}(E)$ and $b_{Ln\alpha(\beta)}^{(\dagger)}(E)$ annihilates (creates) an electron with spin $\alpha$ ($\beta$) and energy $E$ in transverse mode $n$ in the left lead moving towards or leaving the ferromagnet, respectively. The $b$-operators are related to the $a$-operators through the unitary scattering matrix $\mathcal{S}(E)$. Combining all indices $b_{p}(E)=\sum_{q}s_{pq}(E)a_{q}(E)$, where the scattering coefficient $s_{pq}$ is an element of $\mathcal{S}$ and the sum runs over all possible spin states and transverse modes in both the left and the right lead. The charge and spin currents are $I_{c,L}(t)=\sum_{\alpha}\hat{I}_L^{\alpha\alpha}(t)$ and $\mathbf{I}_{s,L}(t)=-\hbar/(2e)\sum_{\alpha\beta}\hat{\mbox{\boldmath $\sigma$}}^{\alpha\beta}\hat{I}_L^{\beta\alpha}(t)$, where $\hat{\mbox{\boldmath $\sigma$}}$ is a vector of the Pauli matrices. With the quantum mechanical expectation value $\langle a_{Ln\alpha}^{\dagger}(E)a_{Lm\beta}(E')\rangle = \delta_{mn}\delta_{\alpha\beta}\delta(E-E')f(E-\mu_L)$, the average charge and spin currents can be obtained \cite{prl84epjb22}. The charge current fluctuations on the left side of the ferromagnet are given by the correlation function $S_{c,LL}(t-t') = \langle\Delta I_{c,L}(t)\Delta I_{c,L}(t')\rangle$, where $\Delta I_{c,L}(t)=I_{c,L}(t)-\langle I_{c,L}(t)\rangle$ is the fluctuation of the charge current from its average value. Expressions are in the following simplified by assuming that the normal metals are either very large or support strong spin-flip scattering, such that a spin current emitted by the ferromagnet never returns. We also assume that the ferromagnet is thicker than the magnetic coherence length. Furthermore, we disregard spin-flip processes in the ferromagnet, which is allowed when the spin-flip length is longer than the coherence length. We assume that the noise frequencies are much smaller than both the temperature, the applied voltage and the exchange splitting in the ferromagnet. This is implicit in Eq. (\ref{h0corr}) and in adiabatic spin pumping theory \cite{prl88}, and is well justified up to ferromagnetic resonance frequencies in the GHz regime. The average magnetization direction is taken to be along the $z$-axis.

Let us consider first the unbiased trilayer with zero average current. At $T \ne 0$ the instant current at time $t$ does not vanish due to thermal fluctuations. The zero frequency thermal charge current noise is found by Fourier transforming the current correlation function. The result is $S_{c,LL}^{(\rm th)}(\omega=0)=2k_BT(e^2/h)(g^{\uparrow}+g^{\downarrow})$, where $g^{\alpha}=M_L-\sum_{mn}|r_{mn\alpha}|^2$ is the dimensionless spin-dependent conductance, $M_L$ the number of transverse modes in the left normal lead, and $r_{mn\alpha}$ the spin-dependent scattering coefficient for electron reflection from mode $n$ to $m$ on the left side. $r_{mn\alpha}$ should be evaluated at the Fermi energy. This is the well-known Johnson-Nyquist noise that relates the dissipative element, i.e., the electric resistance, to the noise, as required by the FDT.

More interesting is the correlation
\be
	S_{ij,KK'}(t-t') = \langle\Delta I_{i,K}(t)\Delta I_{j,K'}(t')\rangle
\label{corrfunc}
\ee
between the $i$-component ($i=x,y$ or $z$) of the spin current on side $K$ ($=L$ or $R$) and the $j$-component ($j=x,y$ or $z$) on side $K'$ ($=L$ or $R$). The zero frequency thermal spin current noise becomes:
\be
\label{eqspinnoise}	 
	S_{ij,KK'}^{(\rm th)} = \frac{\hbar k_BT}{8\pi}\sum_{\alpha\beta}\hat\sigma^{\alpha\beta}_{i}\hat\sigma^{\beta\alpha}_{j} 
				(2M_K\delta_{KK'}-Q_{KK'}^{\alpha\beta}-Q_{K'K}^{\beta\alpha}),
\ee
where $\sigma_i$ denotes one of the Pauli matrices ($i=x,y,z$), $M_K$ is the number of transverse modes in lead $K$, and $Q_{KK'}^{\alpha\beta}={\rm Tr}(s_{KK'\alpha}^{\dagger}s_{KK'\beta})$ should be evaluated at the Fermi energy. The trace is taken over the space of the transverse modes that span the matrices of the scattering coefficients. The $xx$- and $yy$-components of the thermal spin current noise, $S_{xx,LL}^{(\rm th)}$ and $S_{yy,LL}^{(\rm th)}$, are governed by the real part of the dimensionless mixing conductance \cite{prl84epjb22} $g_L^{\uparrow\downarrow}=M_L-\sum_{mn}r_{mn\uparrow}r_{mn\downarrow}^*$. Furthermore, $S_{xx,LL}^{(\rm th)} \neq S_{xx,LR}^{(\rm th)}$ (and similar for the $yy$-component) since the transverse spin current is not conserved at the interface. By angular momentum conservation absorption of the fluctuating spin current implies random torques acting on the magnetization. On the other hand (in the absence of spin flip scattering) $S_{zz,LL}^{(\rm th)} = S_{zz,LR}^{(\rm th)}$ since a spin current polarized parallel to the magnetization is allowed to traverse the ferromagnet.

We now turn to the effect of the fluctuating torques on the magnetization vector. To this end the LLG Eq. (\ref{LLG}) must be generalized by substituting $d\mathbf{m}/dt \rightarrow d\mathbf{m}/dt+ \gamma\mathbf{I}_{s,\rm abs}/(M_s\mathcal{V})$, where $M_s\mathcal{V}$ is the total magnetization of the ferromagnet and $\mathbf{I}_{s,\rm abs} = \mathbf{I}_{s,L}-\mathbf{I}_{s,R}$ is the spin current absorbed by the ferromagnet. The mean $\langle \mathbf{I}_{s,\rm abs} \rangle$ vanishes for the single ferromagnet considered here, but the fluctuations $\langle (\mathbf{I}_{s,\rm abs})^2 \rangle $ do not. The thermal magnetization noise of the isolated magnet is given by Eq. (\ref{h0corr}). Proceeding from Eq. (\ref{eqspinnoise}) we find the thermal fluctuations of the torque to be of exactly the same form and therefore represented by a new, statistically independent random field $\mathbf{h}^{(\rm th)}(t)$ with correlation function 
\be	
	\langle h_i^{(\rm th)}(t)h_j^{(\rm th)}(t') \rangle = 2k_BT\frac{\alpha'}{\gamma M_s\mathcal{V}}\delta_{ij}\delta(t-t'),
\label{eqnoise}
\ee
where $\alpha'$ is defined by
\be	
	\alpha' = \frac{\gamma\hbar {\rm Re}(g_{L}^{\uparrow\downarrow}+g_R^{\uparrow\downarrow})}{4\pi M_s\mathcal{V}},
\label{alpha'}
\ee
and where $i$ and $j$ label axes perpendicular to the magnetization direction. The condition that the ferromagnet is thicker than the coherence length allowed us to disregard terms like $\sum_{mn}t_{mn\uparrow}t_{mn\downarrow}^*$, where $t_{mn\uparrow(\downarrow)}$ is the spin-dependent transmission coefficient for electron propagation from the left side to the right. The expression for $\alpha'$ is identical to the enhancement of the Gilbert damping in adiabatic spin-pumping theory \cite{prl88}.  We conclude that the enhanced magnetization noise in N|F|N sandwiches can be described by an effective random field $\mathbf{h}(t)=\mathbf{h}^{(0)}(t)+\mathbf{h}^{(\rm th)}(t)$, associated with the enhanced Gilbert constant $\alpha=\alpha_0+\alpha'$. Basically, we extended the LLG with a (Langevin) thermal agitation term given by $\mathbf{h}^{(\rm th)}(t)$ to capture the increased noise that, according to the FDT, must exist in the presence of spin pumping. We proved that the thermal spin-current noise is the underlying microscopic mechanism. Large magnetization noise is expected in thin magnetic layers in which $\alpha'$ dominates $\alpha_0$ \cite{Tserkovnyakreview}. The small imaginary part of the mixing conductance does not appear explicitly in Eq. (\ref{alpha'}). Via a renormalized gyromagnetic ratio $\gamma$ \cite{prl88} it affects $\mathbf{h}^{(0)}(t)$ and $\mathbf{h}^{(\rm th)}(t)$ identically, keeping the FDT intact.

The shot noise is most easily evaluated at zero temperature. Using Eq. (\ref{corrfunc}) we find $S_{c,LL}^{(\rm sh)}=S_{c,LR}^{(\rm sh)}$, reflecting charge conservation. The zero-frequency spin shot noise at $T=0$ is
\bea
\label{shspinnoise}
	S_{ij,KK'}^{(\rm sh)} & = & \frac{\hbar}{8\pi}\sum_{\alpha\beta}\hat\sigma^{\alpha\beta}_{i}\hat\sigma^{\beta\alpha}_{j}\int dE  \\		 			&& \sum_{K''\neq K'''}W_{KK'K''K'''}^{\alpha\beta}f_{K'''}(1-f_{K''}), \nonumber
\eea
where $i,j=x$ or $y$, $K'',K'''=L$ or $R$, and $W_{KK'K''K'''}^{\alpha\beta}={\rm Tr}[s_{KK'''\alpha}^{\dagger}s_{KK''\beta}s_{K'K''\beta}^{\dagger}s_{K'K'''\alpha}]$. Non-conservation of the transverse spin shot noise implies a fluctuating torque as above. Using Eq. (\ref{shspinnoise}) we obtain the magnetization noise induced by the spin shot noise, 
\bea
\label{shotnoise}	
	\langle h_i^{(\rm sh)}(t)h_j^{(\rm sh)}(t') \rangle & = & \frac{\hbar}{4\pi}\frac{e|V|}{M_s^2\mathcal{V}^2}\delta_{ij}\delta(t-t') \\		 
							&& \times [{\rm Tr}(r_{\uparrow}r_{\uparrow}^{\dagger}t'_{\downarrow}t_{\downarrow}^{\prime\dagger})+
								{\rm Tr}(r'_{\downarrow}r_{\downarrow}^{\prime\dagger}t_{\uparrow}t_{\uparrow}^{\dagger})], \nonumber
\eea
where $\mu_L-\mu_R=eV$ is the applied voltage and $r$ $(r')$ and $t$ $(t')$ are the reflection and transmission matrices for electron propagation from left (right) to right (left). A number of terms in the second sum in Eq. (\ref{shspinnoise}) have been disregarded using the condition that the ferromagnet is thicker than the coherence length. Eq. (\ref{shotnoise}) vanishes with the exchange splitting only if these terms are included.

In order to compare the shot noise, Eq. (\ref{shotnoise}), with the thermal noise, Eq. (\ref{eqnoise}), we consider a symmetric N|F|N structure (Fig. \ref{system}) with clean interfaces that conserve the transverse momentum of scattering electrons. We adopt a simple semiclassical approximation in which an incoming electron is totally reflected when its kinetic energy is lower than the potential barrier of the ferromagnet, and transmitted with unit probability otherwise. In terms of the exchange splitting $\Delta U=U_{\uparrow}-U_{\downarrow}$, where $U_{\uparrow(\downarrow)}$ is the potential barrier for spin-up (down) electrons, the combination of scattering coefficients is simplified to
\be	
	{\rm Tr}(r_{\uparrow}r_{\uparrow}^{\dagger}t'_{\downarrow}t_{\downarrow}^{\prime\dagger})+
				{\rm Tr}(r'_{\downarrow}r_{\downarrow}^{\prime\dagger}t_{\uparrow}t_{\uparrow}^{\dagger}) = M\frac{\Delta U}{E_F},
\ee
where $M$ is the number of transverse modes and $E_F$ the Fermi energy in the normal metal. With $\sum_{mn}r_{mn\uparrow}r_{mn\downarrow}^*\approx 0$, which usually holds for intermetallic interfaces, the mixing conductance reduces to $g_{L}^{\uparrow\downarrow}=g_R^{\uparrow\downarrow}=M$. The condition for a significant contribution of shot noise to the magnetization noise can thus be written $eV>k_BTE_F/\Delta U$. For $\Delta U\sim E_F/5$ and typical experimental voltage drops in nanoscale spin valves this condition is $T\lesssim 10\rm K$. At low temperatures we therefore predict an observable crossover from thermal to shot noise dominated magnetization noise as a function of the applied bias.

The effective random field $\mathbf{h}(t)$ is not directly observable, but its correlation function is readily translated into that of the magnetization vector $\mathbf{m}(t)$. Linearizing the LLG-equation (including spin pumping) in terms of small deviations $\Delta \mathbf{m}$ from the equilibrium direction $\hat z$ we obtain the power spectrum of the $x$-component of the magnetization vector $S_x(\omega) = \int d(t-t')e^{i\omega(t-t')}\langle \Delta m_x(t)\Delta m_x(t') \rangle$,
\be
\label{Smxmx}	
	S_x(\omega) = 2\gamma\frac{\alpha k_BT}{M_s\mathcal{V}}\frac{\omega^2+\omega_y^2+\alpha^2\omega^2}{(\omega^2-\omega_0^2+\alpha^2\omega^2)^2
					+\alpha^2\omega^2(\omega_x+\omega_y)^2},
\ee
and similarly for the $y$-component. Here shot noise has been disregarded, $\alpha$ is the spin-pumping-enhanced Gilbert constant, $\omega_0=\sqrt{\omega_x\omega_y}$ is the ferromagnetic resonance frequency, and $\omega_x$ and $\omega_y$ are determined by the leading terms in the magnetic free energy expansion near equilibrium, where $x$ and $y$ are taken along the principal axes transverse to $z$. Note that Eq. (\ref{Smxmx}) is proportional to the imaginary (dissipative) part of the transverse spin susceptibility in accordance with the FDT. It therefore reflects both the enhanced broadening of the ferromagnetic resonance as well as the enhanced low frequency magnetization noise. Including shot noise increases the prefactor of Eq. (\ref{Smxmx}) with a bias dependent term.

Rebei and Simionato recently investigated magnetization noise in ferromagnetic thin films using an sd-model \cite{Rebei}, and found results similar to our Eq (\ref{Smxmx}). We believe that our approach based on the scattering theory of transport is more general and, not being based on a specific model for the electronic structure, accessible to first-principles calculations \cite{Xia:prb02}, and better suited to treat more complicated devices. Also, Rebei and Simionato did not attempt to evaluate the shot noise contribution to the magnetization noise.

All results are based on the white noise assumption that is justified as long as typical noise frequencies $\hbar\omega \ll k_B T$. After Fourier transforming the correlators the expressions can easily be generalized to hold for finite frequencies by replacing $k_B T \rightarrow (\hbar\omega/2)\coth[\hbar\omega/(2k_B T)]$.

In conclusion, we demonstrate that the magnetization noise in nanoscale ferromagnets is increased by contacting with a conducting environment. The effect is explained by the transfer of transverse spin current fluctuations in the normal conductors to the ferromagnetic order parameter. Both thermal and shot noise generate effective random magnetic fields felt by the magnetization. The thermal magnetization noise increases in the same way as the Gilbert damping of the mean-field magnetization dynamics, in accordance with the fluctuation-dissipation theorem. Just like the spin-pumping induced broadening of the ferromagnetic resonance, the low-frequency magnetization noise is strongly enhanced in thin ferromagnetic films covered by a few monolayers of a strong spin-flip scattering metal such as Pt. At easily accessible lower temperatures the effect of shot noise dominates that of thermal noise. 

We thank Adnan Rebei for discussions and for preprints of unpublished work. This work was supported in part by the Research Council of Norway, NANOMAT Grants No. 158518/143 and 158547/431, and the FOM.

\end{document}